# Multistep Automated Data Labelling Procedure (MADLaP) for Thyroid Nodules on Ultrasound: An Artificial Intelligence Approach for Automating Image Annotation

**Author information**

Corresponding Author:

Jikai Zhang (jikai.zhang@duke.edu)

Department of Electrical and Computer Engineering, Duke University, Durham, NC, United States

Room 10070, 2424 Erwin Rd, Durham, NC, 27705

Maciej A. Mazurowski

Department of Radiology, Duke University Medical Center, Durham, NC, United States

Department of Electrical and Computer Engineering, Department of Biostatistics and Bioinformatics, Department of Computer Science, Duke University, Durham, NC, United States

Room 9044, 2424 Erwin Rd, Durham, NC, 27705

Brian C. Allen

Department of Radiology, Duke University Medical Center, Durham, NC, United States

Duke University, Dept of Radiology, Box 3808, Durham, NC, 27710

Benjamin Wildman-Tobriner

Department of Radiology, Duke University Medical Center, Durham, NC, United States

Duke University, Dept of Radiology, Box 3808, Durham, NC, 27710

**Abstract**

Machine learning (ML) for diagnosis of thyroid nodules on ultrasound is an active area of research. However, ML tools require large, well-labelled datasets, the curation of which is time-consuming and labor-intensive. The purpose of our study was to develop and test a deep-learning-based tool to facilitate and automate the data annotation process for thyroid nodules; we named our tool Multistep Automated Data Labelling Procedure (MADLaP). MADLaP was designed to take multiple inputs included pathology reports, ultrasound images, and radiology reports. Using multiple step-wise 'modules' including rule-based natural language processing, deep-learning-based imaging segmentation, and optical character recognition, MADLaP automatically identified images of a specific thyroid nodule and correctly assigned a pathology label. The model was developed using a training set of 378 patients across our health system and tested on a separate set of 93 patients. Ground truths for both sets were selected by an experienced radiologist. Performance metrics including yield (how many labeled images the model produced) and accuracy (percentage correct) were measured using the test set. MADLaP achieved a yield of 63% and an accuracy of 83%. The yield progressively increased as the input data moved through each module, while accuracy peaked part way through. Error analysis showed that inputs from certain examination sites had lower accuracy (40%) than the other sites (90%, 100%). MADLaP successfully created curated datasets of labeled ultrasound images of thyroid nodules. While accurate, the relatively suboptimal yield of MADLaP exposed some challenges when trying to automatically label radiology images from heterogeneous sources. The complex task of image curation and annotation could be automated, allowing for enrichment of larger datasets for use in machine learning development.

**Keywords:** Data Processing, EHR, Image Segmentation, NLP, OCR

1. Introduction

Machine learning (ML) applications in radiology continue to expand, with numerous algorithms developed for a variety of tasks including study protocoling [1], report analysis [2], and image analysis [3–8]. Models for these tools may be trained using several methods including supervised learning, unsupervised learning, and reinforcement learning. In supervised learning, input data must be accurately labeled, often necessitating manual annotation from medical experts. This requirement poses a significant challenge, as expert time is limited and the process of labeling and organizing data may be significant.

ML for diagnosis of thyroid nodules on ultrasound (US) is an active area of research [9–12] and is representative of the aforementioned challenges that arise during data curation. Creating a labeled dataset of US images for the task of differentiating benign from malignant nodules requires the annotator (often a radiologist) must label US images to reflect pathology results. To do this, the radiologist performs several tasks: he or she must extract information from a pathology report, correlate that data across dozens of images within the US examination, and often further cross reference with information in the radiology report. This series of tasks is complex, and correctly identifying images of a specific thyroid nodule (often among many) as it relates to specific pathology is time-consuming and labor-intensive.

One solution to streamline medical data curation and annotation is to develop automated labeling applications. Indeed, ML-based tools to supplement or even replace manual efforts have been reported. Natural language processing (NLP) algorithms have been applied to radiology reports for document classification [13–17] and clinical entity recognition [18]. Models for annotating segmentation labels in medical images have also been described [19–23]. However, despite high performance, these tools often focus on a single data type, either free text or images, not both. Nor do currently described tools actively link information between text and imaging data in order to curate a fully-annotated dataset from raw data.

With these challenges in mind, the purpose of our study was to create a multistep automated data labelling procedure (MADLaP) for thyroid nodules on US. The goal of MADLaP was to automatically link pathology data to images of specific nodules and to produce a set of US nodule images with corresponding pathology labels. We hypothesized that a combination of existing tools such as natural language processing, optical character recognition (OCR), and deep learning image segmentation could be applied to pathology reports, radiology reports, and US images in order to perform the series of complex tasks normally assigned to a human. Ultimately, an accurate and efficient MADLaP could both produce larger labeled datasets for future thyroid ML efforts and also serve as a framework for other DL data curation tasks outside of thyroid US. Our code is available on GitHub: https://github.com/MaciejMazurowski/MADLaP.

## 2. Methods

### 2.1. Study Population

This retrospective study was HIPAA-compliant and approved by the institutional review board (IRB) of Duke University Medical Center. A waiver of informed consent was granted by the IRB as part of the approval. The initial population included 4,354 patients, identified using electronic health record database of Duke University. Specifically, we searched cytopathology reports over a six-year period (October 2014 – October 2020) using a set of thyroid-specific keywords, including 1) "thyroid, nodule, benign", 2) "thyroid, nodule, carcinoma." These terms allowed for identification of patients with either benign (Bethesda II), suspicious (Bethesda V), or malignant (Bethesda VI) thyroid nodules based on fine needle aspiration (FNA). Intermediate FNA results (Bethesda III and IV) were not included because of the desire for definitive pathology for future algorithm development.

Next, each patient was queried using our institution's picture archiving and communications system (PACS) to identify any US examinations performed within 6 months of the pathology report date. This window was chosen in order to maximize the chances of capturing any thyroid US or thyroid FNA related to the pathology results. At our institution, thyroid US examinations and FNAs are performed by several providers including diagnostic radiology at our main academic center, endocrinology, otolaryngology, and either of our two community hospitals. Patients with imaging from any of these practice environments were included, while patients with no US examinations were excluded. From the above process, we obtained 3,981 patients who had US imaging within six months of their pathology report date to be included. Ultimately, each included patient had three data components: 1) pathology report, 2) US examination (images), and 3) corresponding radiology report (also available through PACS). The final dataset was created by randomly assigning 402 patients to a training set and 100 patients to a test set, and then excluding patients who 1) were <18 years-old, and 2) had no thyroid US imaging (whether from a diagnostic study or images captured during FNA) from any of the aforementioned practice environments.

The final training and test sets consisted of 378 patients and 93 patients respectively, with pathology reports, US studies, and corresponding radiology reports included. In addition, the practice location of the FNA was recorded, including Site 1: our large academic medical center's radiology department, Site 2: outpatient endocrine center, Site 3: otolaryngology, and either of our two community hospitals.

### 2.2. Establishment of Ground-truth

To establish a ground truth for matching nodule images and corresponding pathology, all patients in both the training and test sets were reviewed by a fellowship-trained radiologist with daily practice in thyroid US (B.W.T., 3 years post-fellowship experience). The radiologist reviewed pathology reports in combination with radiology reports and US images to identify US images that corresponded to a nodule that underwent FNA. The radiologist selected two key images: transverse and longitudinal views of every nodule in each patient's pathology report. The radiologist was blinded to the results from the proposed MADLaP during the identification process.

Additionally, a second reader (B.C.A., 10 years post-fellowship experience) provided annotations for a random sample of 30 patients from the test set. These selections were compared to the primary reader's annotations to validate the ground truth selections by the primary reader.

### 2.3. MADLaP Design

The goal of MADLaP was to take multiple inputs (pathology report, US images, radiology report) and automatically identify two key US images of a specific thyroid nodule, and to correctly assign a pathology label. To achieve this goal, MADLaP utilized a combination of existing tools across multiple fields in AI domain, including rule-based NLP models, a pre-trained deep-learning-based imaging segmentation model, and an OCR tool. Fig. 1 shows a complete diagram of MADLaP, with more detailed description of individual components as follows.

MADLaP was developed using the training set. The data at the start of MADLaP included a pathology report and US images; however, what US images were available would dictate MADLaP workflow based on two overall MADLaP stages. For example, if images from a thyroid FNA within the department of radiology were available, MADLaP would start in stage 1. This design reflected the fact that images from thyroid FNAs were more focused examinations and typically depicted only the nodule undergoing FNA (i.e., the desired nodule for labelling), and that US images from the department of radiology were typically well labeled. If no such study was available (i.e., the patient underwent FNA in another department), MADLaP would start in stage 2, which focused on diagnostic thyroid US, studies that typically included more images and possibly more nodules than what had undergone FNA.

Stage 1 consisted of four individual steps (Modules 1-4). Each module performed a specific information extraction task in order to help match US images to pathology results.

Module 1 paired the input pathology report with a corresponding thyroid US study by comparing dates. We considered a pathology report and US study to be a match if report dates were within 6 months of each other. If more than one US had been performed within six months, the study in closest temporal proximity to the pathology report date was selected.

Module 2 focused on pathology reports and consisted of a rule-based NLP algorithm (developed internally, examples shown in Fig. 2) to extract text data about individual nodules from each report. Structured data outputs from the algorithm included laterality (left, right, isthmus), location (superior, anterior, posterior, mid, etc.), and nodule label (#1, #2, etc.). The NLP algorithm first searched for the line of the nodule information, and then searched for the diagnosis in the subsequent lines (Fig. 2). Nodules with a diagnosis matched to thyroid-specific keywords were recorded. Throughout training, pattern-matching pieces were added to the main algorithm to improve the correctness of output nodule information.

Module 3 was then applied to US images and employed a pre-trained deep-learning-based caliper detection model [24] to identify the presence or absence of calipers in every image (Fig. 3). Calipers (measurement annotations made by sonographers and embedded in images) are almost always applied to images of nodules, and this fact was used to narrow the search for key images. After the initial detection phase, images containing exactly 2 or 4 calipers (suggesting a measurement of a nodule) were saved. Since stage 1 of MADLaP consisted of thyroid FNA studies only, caliper detection on extraneous structures in the thyroid (for example a lobe or another structure) were not present. Thus, if caliper detection identified exactly two images containing calipers, and only one nodule was extracted from the pathology report, this was considered a successful match of pathology data to imaging data. If more than two images were selected, images were pushed to the next module within Stage 1 for further refinement. If less than two images were selected, then the pathology report (and extracted nodule data) was passed to the stage 2 for comparison to the full diagnostic US.

For Module 4, we applied an OCR tool to detect text within US images that contained nodule information. To do this, we first cropped the image to include only the bottom of the image, where nodule information is typically annotated. Then, we used the Tesseract-OCR engine [25] to detect and extract text. Various cropping ratios were tested to optimize performance. Based on the OCR output from each image, we developed and applied a rule-based NLP algorithm to record laterality, location, and nodule labels (Fig. 4). Finally, we compared the OCR results with nodule information from the pathology report. If there were two images that contained matching nodule information, we considered a successful match. Otherwise, the pathology report moved on to stage 2.

Stage 2 also consisted of four individual modules. Modules 1, 3, and 4 of stage 2 were similar to corresponding modules in stage 1, but were applied to full diagnostic US examinations instead of focused FNA studies. Briefly, MADLaP matched the correct diagnostic study to a pathology report, applied the caliper detection model to find candidate diagnostic images with 2 or 4 calipers, applied the OCR tool to detect nodule information from images, and matched the OCR outputs with the nodule information extracted from the pathology report. Additional matches were identified at this point, but if there were more than two images selected after Module 4 (i.e., not an exact match), images were sent to a new module, Module 5.

Module 5 was designed to match nodule measurements included in images with measurements in radiology reports. This module was specifically designed in stage 2 because complete radiology reports accompanied diagnostic studies rather than FNA examinations in stage 1. This step could further clarify the output of Module 4 in cases where too many (i.e., more than 2) images were selected for a given nodule. By matching measurement data embedded within US images with data in the radiology report, an exact match could be made. To do this, we first applied the Tesseract-OCR to detect measurements from the images (Fig. 5). Next, we developed and applied a rule-based NLP algorithm to detect nodule measurements within the radiology report (Fig. 6). The NLP algorithm targeted nodule measurement data based on laterality from the pathology report. To ensure precise matching, we applied these steps only for examinations with single thyroid nodules on the side (right/left) in question. Radiology reports with multiple nodules on the side in question were not analyzed.

### 2.4. Evaluation of MADLaP Performance

MADLaP performance was assessed using two metrics: yield rate and accuracy. The raw MADLaP output (i.e., the number of nodules and corresponding images selected by MADLaP) was considered the yield, and the yield rate was defined as MADLaP yield divided by the number of nodules in the test set.

$$YR = \frac{\text{\# of identified nodules (correct matches + incorrect matches)}}{\text{\# of nodules (ground truth)}}$$

Accuracy was determined by directly comparing MADLaP output images to radiologist-established ground truth images. Manual comparison allowed for sorting of MADLaP yield into one of two categories. Correct yields were considered Category 1 (C1), cases with exactly 2 images for a nodule that were identical to what the radiologist selected. Incorrect yields were considered Category 2 (C2), cases where either nodule information extracted by the NLP algorithm was incorrect, or the selected pair of images was different from what the radiologist selected. Accuracy assessed what percent of image outputs from MADLaP were correct and was defined as:

$$Accuracy = \frac{\text{\# of correct matches}}{\text{\# of identified nodules (correct matches + incorrect matches)}}$$

Next, we evaluated the point along MADLaP (as defined by stage and module) at which image matching finalization occurred. Both correct and incorrect yields were evaluated, and yield rate and accuracy for each stage and module were calculated. For correct yields (C1), we reported YR, counts, and accuracy for Stages 1-2, Modules 3-5. For incorrect yields (C2), we evaluated each of these failure points by site of image acquisition. We also compiled descriptive statistics for yield and accuracy by site.

Lastly, MADLaP yield and accuracy were also evaluated using the subset of test set cases annotated by the second reader. Both the first and second reader's annotations were used as ground truth, and MADLaP performance was tested to see if differences emerged when using different reader annotations.

3. Results

3.1. Training and Test Set Composition

The final training set included 462 nodules from 378 patients, and the final test set included 103 nodules from 93 patients. The prevalence of malignant nodules in the training and test sets was 6.9% and 7.8%, respectively. In the training set, 262 nodules were from the site 1, 111 nodules were from the site 2, and 89 nodules were from the site 3. In the test set, 70 nodules were from the site 1, 20 nodules were from the site 2, and 13 nodules were from the site 3.

3.2. Results

MADLaP application to the test set resulted in a yield of 65 nodules and a yield rate (YR) of 63% (65/103). 54 of the image pairs proposed by MADLaP were correct, for an accuracy of 83% (54/65). When evaluated by 6 clinical sites of examination, MADLaP performed better when analyzing USs performed at site 1, with a YR of 71%, and an accuracy of 90%. Using images from Sites 2 and 3 resulted in lower MADLaP performance, with an accuracy and YR of 55% for site 2, and 31% YR for site 3. Complete results of yield by site are found in Table 1.

Table 1. Number of nodules from ground-truth, number of nodule output from MADLaP, number of successful matches, Yield Rate, and Accuracy from the test set.

| Site # | # Ground-truth nodules | # Yields | # Successful Yields | Yield Rate | Accuracy |
|--------|------------------------|----------|---------------------|------------|----------|
| 1      | 70                     | 50       | 45                  | 71%        | 90%      |
| **2** | 20 | 11 | 5 | 55% | 45% |
| **3** | 13 | 4 | 4 | 31% | 100% |
| **All Sites** | 103 | 65 | 54 | 63% | 83% |

Each MADLaP module was analyzed individually for performance. As expected, the cumulative yield progressively increased as the input data moved through each module. The largest increase in yield occurred at stage 1 M4, where yield increased from 20% to 50%. Yield increased slightly throughout Stage 2, ending at 63%. Accuracy was also highest at Stage 2 M4 (86%), and slightly decreased as data moved forward, ending at 83% as the yield slightly increased. Table 2 shows complete performance statistics from MADLaP by stages and modules.

Table 2. Yields (YR) and number of successful matches (accuracy) at the end of the module (that produced yields) of each stage.

| | Stage 1[*] (search through FNA imaging studies) | | Stage 2[**] (search through diagnostic imaging studies) | |
|---|---|---|---|---|
| | M3 | M4 | M4 | M5 |
| **Yield (YR)** | 21 (20%) | 51 (50%) | 59 (56%) | 65 (63%) |
| **Correct yields (Accuracy)** | 20 (95%) | 44 (86%) | 51 (86%) | 54 (83%) |

*: By design, M1 and M2 of stage 1 were not presented because here MADLaP did not produce yields.
**: By design, MADLaP would not produce yields by the end of M3 in stage 2.

The point along MADLaP at which incorrect yields occurred was then analyzed. Overall, incorrect yields were similarly split between Stage 1 (n=7) and Stage 2 (n=4). By module, Stage 1 M4 and Stage 2 M5 had the highest number of incorrect yields, with 4 and 2 respectively. When evaluated by site, Site 3 had the lowest percentage of incorrect yields (0%), but had a low yield overall (n=4). By comparison, Site 1 had 10% (5/50) of its yield incorrect, while Site 2 had the highest incorrect rate at 60% (6/10). Table 3 highlights incorrect yields separated by stage, module, and site.

Table 3. Counts of incorrect yields separated by stage, module, and site of image acquisition.

| Stage | Module | # Site 1 N=50 | # Site 2 N=10 | # Site 3 N=4 | Module Total | Stage Total |
|---|---|---|---|---|---|---|
| **Stage 1** | M3 | 0 | 1 | 0 | 1 | 7 |
| | M4 | 5 | 1 | 0 | 6 | |
| **Stage 2** | M4 | 0 | 1 | 0 | 1 | 4 |
| | M5 | 0 | 3 | 0 | 3 | |
| **Site Total*** | | 5 (10%) | 6 (60%) | 0 (0%) | | 11 |

* Numbers in parenthesis denote proportion of yield that is incorrect by site.

Using a random sampling of 30 test set patients (including 46 nodules), reader 1 and reader 2 had a 91% agreement on a per-nodule basis. Out of 39 nodules for which reader 1 matched images to pathology results, reader 2 agreed 100% of the time. Using reader 1's annotations from this sample,

MADLaP yield was 58% and accuracy were 91%. For comparison, when reader 2's annotations were used as ground truth, MADLaP yield and accuracy were 51% and 91%, respectively.

## 4. Discussion

Training ML models in radiology often requires manual annotation of large datasets, a time-consuming and impractical process. Our project showed that a Multistep Automated Data Labelling Procedure (MADLaP) could successfully create datasets of US images of thyroid nodules for downstream machine learning tasks. MADLaP leveraged basic imaging and medical record data (image calipers, embedded text in US images, pathology/radiology text reports) to automatically annotate images. The annotation process was accurate, though its yield was lower than anticipated. MADLaP performance varied greatly depending on the location of original image acquisition. Nonetheless, our results suggest that in full automation, MADLaP could be implemented to curate large datasets using unlabeled, raw data from our institutional database.

Automating the complex process of radiology image annotation proved challenging, though results from MADLaP suggest that automation is feasible. Indeed, other works have demonstrated frameworks for analyzing both free texts and images together. Shin et al. proposed a deep learning model to automatically annotate chest X-ray images based on images and reports [26]. Their labeling process was deemed successful, but differed from our task in that only one image (the chest X-ray) was preselected, as opposed to identifying key images among several within an ultrasound examination. The added layer of complexity related to specific image identification within a larger set makes direct comparison challenging. In another example of linking text data to images, Demner-Fushman et al. utilized caption texts and images to annotate and retrieve clinically relevant images in publications, a helpful approach but not applied to a clinical radiology task [27]. We anticipate that more comprehensive automated tools will emerge as ML efforts in radiology continue.

While accurate, the relatively suboptimal yield of MADLaP exposed some challenges when trying to automatically label radiology images. To mimic the intricate process normally performed by humans, MADLaP required multiple steps and several tools, including NLP, OCR, and ML image analysis. Each of those steps was a potential source of error, which when compounded resulted in lower yield. However, the accuracy of MADLaP was reasonably high, such that if applied to enough data, output could be sufficient for model training. In addition to overall yield, MADLaP had significant variability across our different clinical sites. Compared to other imaging locations such as endocrinology or surgery, performance was highest when using images from our radiology department. This result is likely due to consistent image annotation from sonographers in the radiology department as well as consistent radiology reporting from our core group of radiologists. Consequently, the generalizability of MADLaP appears greatly affected by the standardization of MADLaP inputs, an unsurprising result but a critical consideration when designing tools that may be used across multi-site hospital systems or even for multi-center clinical trials.

While our MADLaP showed room for improvement, it also had some design features that were noteworthy when considering an automated labeling system. First, no annotated input data were required, allowing for full automation based simply on patients coming to the hospital for thyroid FNA. Second, MADLaP was trained on a relatively small dataset, creating a pipeline without requiring overly burdensome input by a radiologist. Had it our primary reader had to annotate a large number of baseline annotations, the utility of the algorithm would be diminished, as a well annotated dataset would already be available. Third, in addition to the primary labelling goal of benign or malignant, ancillary information was automatically collected. In our case, information such as nodule laterality, location, and label may be used for additional research questions. Third, the sequential design of MADLaP allowed for evaluation of individual modules. As such, specific modules could be targeted for improvement, which might both improve overall performance and allow for assessment of tradeoff between yield and accuracy. For example, implementation of M5 resulted in an increase in yield but a loss in accuracy, suggesting that initial attempts to improve MADLaP performance could start with this step. The M5 task of information extraction from radiology reports using NLP is a complex problem for which additional efforts are likely warranted. Ultimately, general guidelines for automated data curation in radiology aided by ML do exist [28, 29], and some of the aforementioned aspects of our MADLaP reflect suggestions within these guidelines.

The proposed MADLaP had several limitations. First, MADLaP was designed on data from a single institution. The significant heterogeneity across sites within our own health system suggests that increasing generalizability of such a MADLaP may be challenging. Additional work, including modifying NLP algorithms and adjusting cropping rules for OCR imaging inputs, may form the basis for improvement. Second, datasets created by our algorithm are intended to be useful for training downstream deep learning models for classification. However, the output accuracy may still be insufficient if an AI task requires near perfect labelling. Future efforts could focus on revising rules in NLP algorithms, utilizing other pre-trained segmentation models for caliper detection, and improving OCR algorithms to improve performance. Third, our training and test sets were annotated by a single reader, which has the potential to introduce bias into the model. However, to evaluate potential bias, results from the second reader's annotations of a random test set sample showed good agreement of image selections with the primary reader. Moreover, when using that same subset of cases, MADLaP accuracy did not change regardless of which reader's annotations were used as ground truth. Lastly, MADLaP was not able to process "double-view" images, those with both transverse and longitudinal views included in a single frame, an additional and more specific example of limited generalizability. However, with additional training, processing of these images could likely be achieved.

5. Conclusion

In this work, we developed and evaluated a multistep automated data labeling procedure (MADLaP) for thyroid nodules on US that could facilitate curation of large datasets for machine learning purposes. Such a tool may also serve as an example for other disease processes in ongoing efforts to reduce the burden of manual labeling.

**DATA AVAILABILITY STATEMENT**

Portions of data may be shared on reasonable request to the corresponding author, though IRB restrictions may limit sharing in some instances.

**Funding**

This work was supported in part by AI SPARK Award from Duke Center for Artificial Intelligence in Radiology.

**FIGURES:**

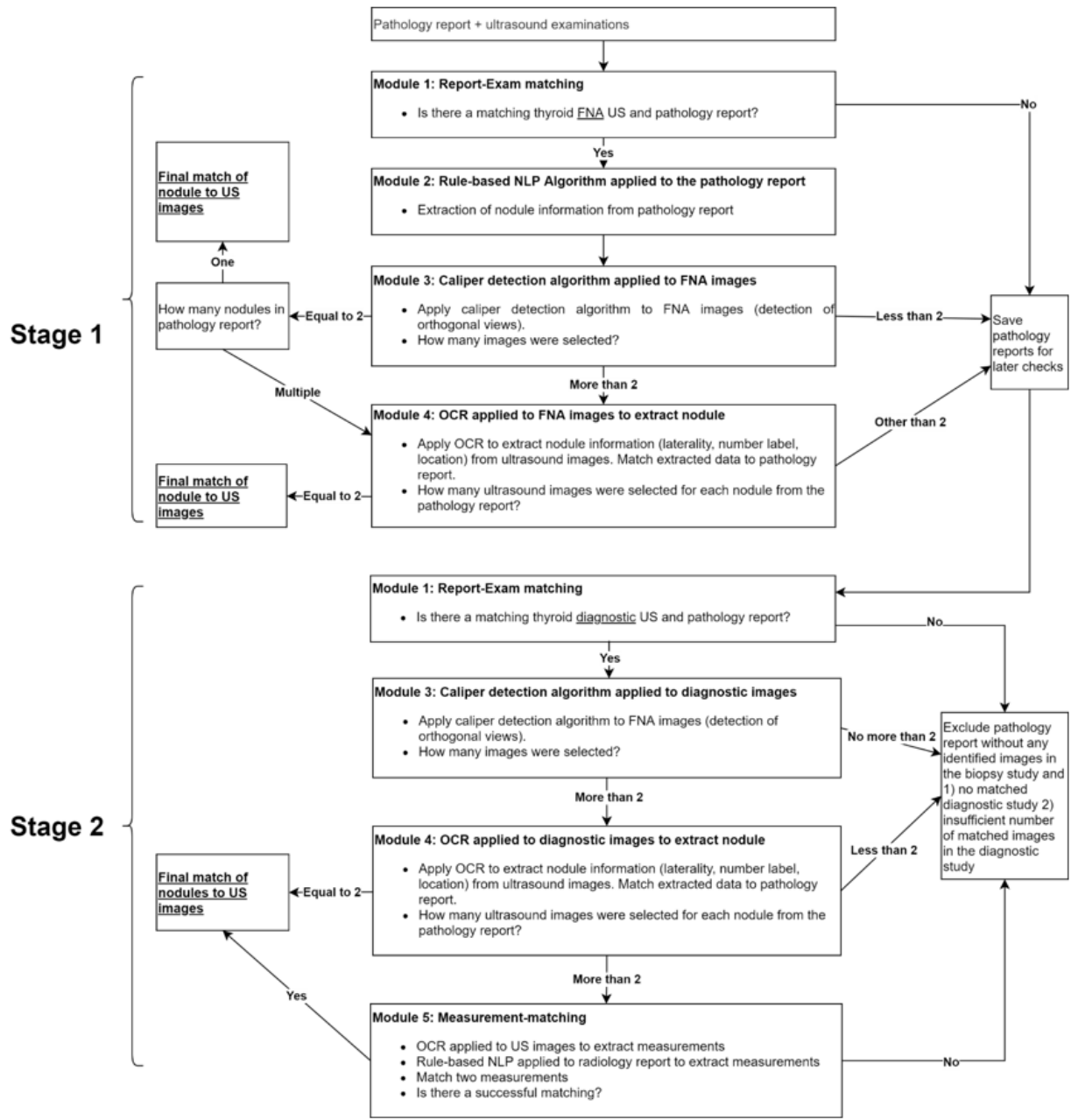

Fig. 1 Flowchart Diagram of MADLaP.

Example 1: [pathology report image with yellow, green, and red highlights]

Example 2: [pathology report image with yellow, green, and red highlights]

Fig. 2 Two examples (a) and (b) of NLP algorithm for processing pathology reports. (a) The algorithm first identified and then checked lines in the "Specimen" section (yellow highlight). The algorithm found the word "Thyroid" (a keyword for locating a thyroid nodule). The algorithm then searched for a 'diagnosis' line (green highlight). Once targeted, the algorithm searched the following line for a final diagnosis (red highlight). This example was a malignant case, the algorithm recorded "right" as laterality, "inferior" as the location, and "#1" as the nodule label. (b) In this example, the Specimen section (yellow highlight) is again identified, but this time with two nodules. The subsequent "Diagnosis" sections (green highlights) reflect two corresponding nodules, which are extracted in order. Each nodule was accompanied with a benign diagnosis (red highlight). In the output, the algorithm recorded "isthmus" as laterality for the first nodule, and "right" for the second nodule. No location or nodule label was found (as none was present).

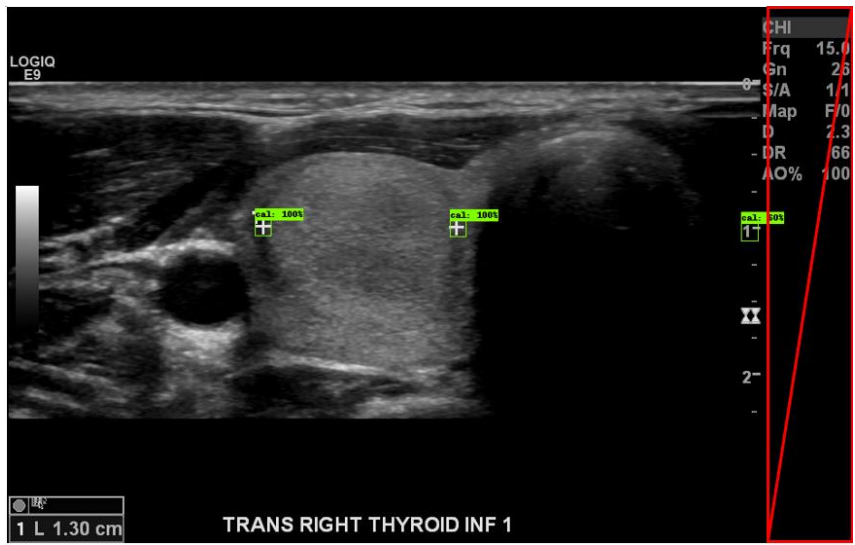

Fig. 3 Automated caliper detection. The detection algorithm segments areas of interest with potential caliper shapes, and each area has a corresponding probability score for being a 'true' caliper. In this case, three areas were selected: two had a score of 100%, and one a score of 60%. A cropping ratio (87%) was applied to crop out right side of the image to avoid unexpected calipers (see red area). A threshold (94.5%) was applied for filtering caliper scores. Thus, the image had two calipers correctly identified.

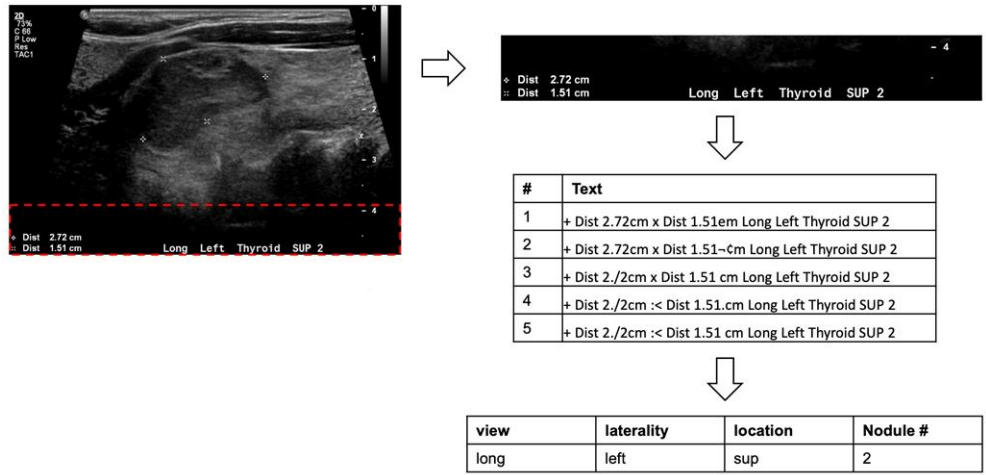

Fig. 4 OCR processing for nodule information. First, images were first automatically cropped (red-dashed box) to focus the area of OCR input. Second, an OCR engine with five parameter configurations extracted text. Third, an NLP algorithm was applied to the give configurations to record final nodule information. In this example, text included view (transverse), laterality: right, location: mid, and nodule label: #1.

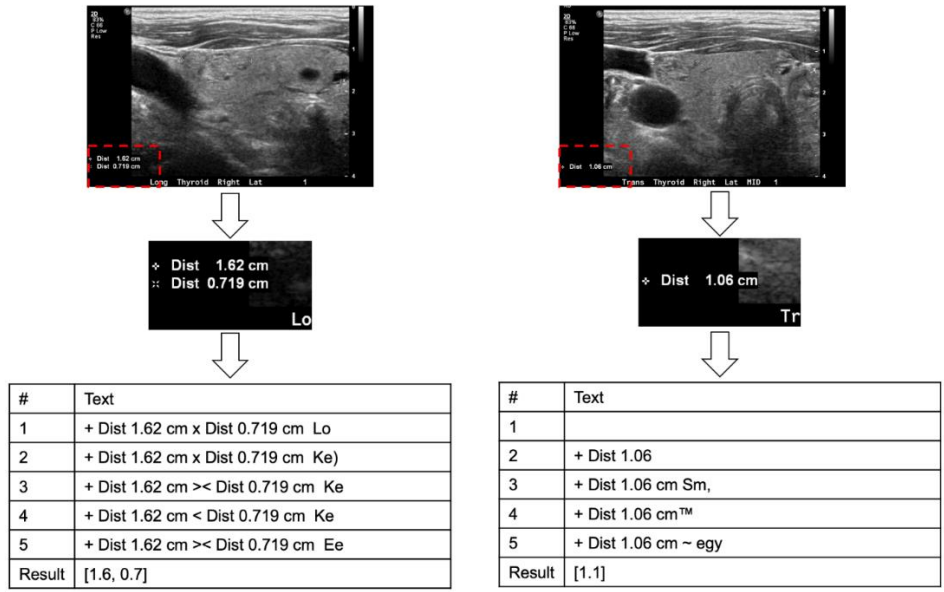

Fig. 5 OCR processing for measurement extraction from images. An OCR engine with five parameter configurations was applied to extract text from a cropped region, and an NLP algorithm was then applied to extract the measurements. In this example, final output measurements match the embedded text distances.

```
Rpt=US soft tissue head and neck,MRN=        ,Date=         ,Facility=    , Acc Num=
Indication: E04.1 Nontoxic single thyroid nodule Large Thyroid Nodule on CT
Comparison: CT neck
Technique: Gray-scale and color Doppler images of the thyroid gland were
obtained.
```
==Findings:==
==RIGHT LOBE:==
The right thyroid lobe is homogeneous in background echotexture. The right
lobe measures 4.2 x 2.1 x 2.1 cm. One nodule is noted:
*  ==Right nodule #1:== Ovoid hyperechoic solid nodule in the right
superolateral thyroid with ill-defined borders without echogenic foci. The
nodule measures ==1.6 x 0.7 x 1.1 cm==.
LEFT LOBE:
The left thyroid lobe is markedly enlarged, with mass effect on the
trachea, and predominately consists of a large left nodule. The left lobe
measures approximately 5.9 x 9.0 x 4.4 cm. One nodule:
*  Left nodule #1: Large heterogeneously isoechoic to hyperechoic solid
nodule with possible septations. The nodule measures up to 7.0 x 4.1 x 5.1
cm.
ISTHMUS:
The isthmus measures 0.4 cm.
Impression:
1. Large heterogeneous left thyroid nodule. Recommend correlation with
reported prior fine-needle aspiration. Suggest comparison with outside
hospital ultrasound images if available to evaluate for change in size (no
ultrasound comparison available at the time of this report).
2. Mass effect on the trachea by the left thyroid as seen on prior CT.
3. Right thyroid nodule as above. Suggest follow-up ultrasound in one year
if this could affect clinical management.
Electronically Signed by:
Electronically Signed on:

Fig. 6 Measurement extraction from radiology reports using NLP. First, the algorithm located the "Findings" section (yellow highlight). Then, laterality was identified using information acquired earlier in MADLaP, in this instance the right side (green highlight). Next, the algorithm searched for nodule-specific language (blue highlight). Finally, the algorithm extracted measurement data (red highlight). To improve accuracy, the algorithm made a final check that the noun before "measures" was "nodule", not "lobe" or other structures. The report measurement was then compared to those extracted from the images.